%% file: main_v5.tex
\documentclass[conference]{IEEEtran}
\IEEEoverridecommandlockouts
\usepackage{cite}
\usepackage[utf8]{inputenc}
\usepackage[T1]{fontenc}
\usepackage{amsmath,amssymb,amsfonts}
\usepackage{graphicx}
\usepackage{subcaption}
\usepackage{textcomp}
\usepackage{xcolor}
\usepackage{booktabs}
\usepackage{array}
\usepackage{url}
\usepackage{multirow}
\usepackage{enumitem}
\usepackage{tikz}
\usetikzlibrary{arrows.meta,positioning,shapes.geometric,fit,calc}
\usepackage{flushend}
\usepackage[hidelinks]{hyperref}
\def\BibTeX{{\rm B\kern-.05em{\sc i\kern-.025em b}\kern-.08em
    T\kern-.1667em\lower.7ex\hbox{E}\kern-.125emX}}

\begin{document}

\title{CYBERFORT: A Compliance-Chain Platform Operationalising the Cyber Resilience Act for SMEs}

\author{%
\IEEEauthorblockN{%
N.~Kekatos\textsuperscript{1},
G.~Koutidis\textsuperscript{1},
M.~Curca\textsuperscript{2},
A.~Fountoulakis\textsuperscript{3},
M.~Ioannou\textsuperscript{4},
M.~Ioannou\textsuperscript{5},\\
E.~Iosif\textsuperscript{6},
P.~Lacatus\textsuperscript{7},
A.~Lekidis\textsuperscript{8},
A.~Michael\textsuperscript{9},
T.~Nianios\textsuperscript{1}%
\thanks{Open-source code (free to use), CRA seed data, framework mappings, and pilot artefacts: \url{https://github.com/CloneSystems/cyberfort_v3}.}}
\IEEEauthorblockA{%
\parbox{0.97\textwidth}{\centering\footnotesize
\textsuperscript{1}Clone Systems, Cyprus, {\ttfamily \{nkekatos, gkoutidis, tnianios\}@clone-systems.com};\
\textsuperscript{2}Directoratul Na\textcommabelow{t}ional de Securitate Cibernetic\u{a}, Romania, {\ttfamily Mihaela.Curca@dnsc.ro};\
\textsuperscript{3}Eximprod Engineering, Romania, {\ttfamily angelos.fountoulakis@epg.ro};\
\textsuperscript{4}Columbia Shipmanagement, Cyprus, {\ttfamily marios.ioannou@csmcy.com};\
\textsuperscript{5}Bolton Technologies, Cyprus, {\ttfamily michael@bts.com.cy};\
\textsuperscript{6}Dealio, Cyprus, {\ttfamily elias.iosif@dealio.ai};\
\textsuperscript{7}I-EnergyLink, Romania, {\ttfamily paul.lacatus@i-energylink.eu};\
\textsuperscript{8}University of Thessaly, Greece, {\ttfamily alekidis@uth.gr};\
\textsuperscript{9}Elias Neocleous \& Co.\ LLC, Cyprus, {\ttfamily Avgi.michael@neo.law}}}}

\maketitle
{\renewcommand{\thefootnote}{}\footnotetext{\copyright~2026 IEEE. Personal use of this material is permitted. Permission from IEEE must be obtained for all other uses, in any current or future media, including reprinting/republishing this material for advertising or promotional purposes, creating new collective works, for resale or redistribution to servers or lists, or reuse of any copyrighted component of this work in other works. This work has been accepted for publication at the 2026 IEEE International Conference on Cyber Security and Resilience (IEEE CSR 2026).}}

\begin{abstract}
The EU Cyber Resilience Act (CRA) turns product cybersecurity into a lifecycle compliance obligation for manufacturers, importers, distributors, and integrators of products with digital elements on the EU market, a load that falls largely on small and medium-sized enterprises (SMEs) that rarely have dedicated governance, risk, and compliance (GRC) capacity. We present CYBERFORT, an open-source CRA-first compliance platform developed under the EU Digital Europe Programme and one of twelve projects in the EU CRA cluster. CYBERFORT operationalises the CRA through a guided scope self-assessment, a question bank tied to Annex~I and the vulnerability-handling obligations, and a compliance-checking engine that links every answer to controls, policies, and machine-attested evidence, reusing ISO/IEC~27001, NIS2, and GDPR controls only where they coincide with CRA obligations. Its central contribution is the \emph{compliance chain}, a traceable structure linking each product risk through its controls and policies to the CRA obligations it satisfies, and onward through evidence to the technical-documentation file and EU declaration of conformity, so that every operational gap is traceable and can be closed before market placement. Deployed at \url{https://access.cyber-fort.eu/} for a first cohort of 43 organisations, the platform is presented with the engineering behind the chain, measured results from a completed end-to-end case study on a SIEM/XDR product with AI-driven remediation spanning the CRA obligation chapters, and the controlled effort study that remains in progress.
\end{abstract}


\section{Introduction}
The Cyber Resilience Act (CRA), Regulation (EU)~2024/2847, establishes horizontal cybersecurity requirements for products with digital elements placed on the EU market \cite{eu_cra_2024}. Obligations span the product lifecycle, from secure-by-design development and risk assessment through vulnerability handling and reporting to technical documentation and EU conformity assessment \cite{eu_cra_2024}. The burden falls disproportionately on SMEs: ENISA notes that SMEs make up the bulk of manufacturers, distributors, importers, and suppliers of products with digital elements in Europe \cite{enisa_sme_cra}, yet only 12.3\% of surveyed SMEs were aware of the CRA in 2024 compared with 83.5\% of very large enterprises \cite{szedlak2024cra}. These SMEs typically run security governance on fragmented spreadsheets, isolated ticketing systems, ad-hoc policies, and disconnected scanners.

CRA compliance is therefore an ongoing operational task, not a one-off legal checklist: the SME must track assets, risks, vulnerabilities, controls, evidence, incidents, and obligations across overlapping regimes (ISO/IEC~27001, NIS2, GDPR), where a single CRA control often reuses controls already implemented for another framework. Without structured mapping, that overlap is rarely captured.

This paper presents CYBERFORT, an open-source multi-framework GRC platform that operationalises CRA requirements into a repeatable workflow. CYBERFORT is developed under the EU Digital Europe Programme (Grant No.~101190281) and is one of twelve EU-funded projects in the CRA cluster. Its central design contribution is the \emph{compliance chain} (Section~\ref{sec:compliance-chain}), which makes every operational gap traceable so it can be assigned, monitored, and closed before market placement. CYBERFORT is deployed at \url{https://access.cyber-fort.eu/} with first-cohort SME tenants.

\subsection{Contributions}\label{sec:contrib}
\begin{itemize}[leftmargin=*]
    \item A \textbf{compliance-chain model} that treats products, risks, controls, policies, CRA obligations (articles and Annex~I requirements), framework mappings, evidence, corrective actions, and conformity documents as connected nodes rather than parallel artefacts.
    \item Four \textbf{CRA-oriented modules} (compliance mapping and gap analysis, risk assessment, control and policy management, and EU conformity) on a multi-framework data model that reuses ISO/IEC~27001, NIS2, and GDPR controls and links one evidence object to many requirements, plus an \textbf{integrated scanner layer} (OWASP~ZAP, Semgrep, Syft, OSV~Scanner, Nmap)\footnote{Scanner tools. OWASP~ZAP: \url{https://www.zaproxy.org/}; Semgrep: \url{https://semgrep.dev/}; Syft: \url{https://github.com/anchore/syft}; OSV-Scanner: \url{https://github.com/google/osv-scanner}; Nmap: \url{https://nmap.org/}} that turns SAST/DAST/SBOM/SCA/network findings into Evidence objects attributable to Annex~I obligations, with a dedicated AI-control family (SCF~v2025.1 AAT~\cite{scf_aat}) for autonomous-operation risks.
    \item A \textbf{deployed, open-source platform} serving 43 organisations, with an engineering account of the chain (Section~\ref{sec:impl}) and measured results from a completed case study on a SIEM/XDR product with AI-driven remediation (Section~\ref{sec:eval}).
\end{itemize}

\section{Background and Related Work}
\paragraph{CRA in brief} The CRA entered into force in December~2024; reporting (Art.~14) applies from September~2026 and the full essential requirements (Annex~I) from December~2027 \cite{ec_cra_policy}. Manufacturers must conduct a risk assessment, embed it into the lifecycle, operate vulnerability handling, and prepare technical documentation and an EU declaration of conformity \cite{eu_cra_2024}. Legal-analytic work frames the CRA as a hybrid regime of command-and-control duties, risk-based calibration, harmonised standards, and enforced self-regulation \cite{teichmann2025clsr}.

\paragraph{SME preparedness} CRA awareness among SMEs is low \cite{szedlak2024cra}, and Conceicao~et~al.\ apply design science to a lightweight mapping linking commonly implemented controls to CRA, ISO/IEC~27001, and CIS for resource-constrained SMEs \cite{conceicao2026sme}. Prior EU SME-security initiatives and open-source CRA-automation work address parts of this problem but rarely deliver an end-to-end, traceable CRA workflow.

\paragraph{EU CRA cluster} CYBERFORT is one of twelve EU-funded CRA-cluster projects, most supported for MSME compliance under the Digital Europe Programme~\cite{ec_cra_msme,enisa_cra_guidance}. Some coordinate harmonised standards and build SME cyber-resilience capacity (CYBERSTAND, STAN4CRA, and SECURE4SME)\footnote{Standards and SME-capacity projects. CYBERSTAND: \url{https://cyberstand.eu/}; STAN4CRA: \url{https://www.stan4cra.eu/}; SECURE4SME: \url{https://www.secure4sme.eu/}}, while most provide open-source tools and conformity solutions for SMEs, spanning SBOM, conformity assessment, compliance automation, AI-driven conformity, IoT and microelectronics certification, an SME compliance wizard, and SaaS certification (OCCTET, OSCRAT, CONFIRMATE, CRACY, CRA-AI, CURIUM, CRACoWi, and TRUSTBOOST)\footnote{Tools- and solutions-for-SME projects. OCCTET: \url{https://occtet.eu/}; OSCRAT: \url{https://oscrat.eu/}; CONFIRMATE: \url{https://confirmate-project.eu/}; CRACY: \url{https://cra-cy.eu/}; CRA-AI: \url{https://www.cybercertlabs.com/cra-ai/}; CURIUM: \url{https://curium-project.eu/}; CRACoWi: \url{https://www.cracowi.eu/}; TRUSTBOOST: \url{https://www.trustboost.eu/}}. CYBERFORT complements these standards- and tool-focused efforts by emphasising an end-to-end integration layer, the compliance chain, which is not the primary deliverable of the other cluster projects.

\paragraph{Multi-framework GRC market} Commercial GRC tooling broadly falls into enterprise suites, compliance-automation tools, and supplier-rating, risk-quantification, and ecosystem-native suites, each covering only part of the CRA lifecycle: enterprise suites are organisation-centric and high-cost, with no product scope, criticality class, operator role, or declaration drafting; attestation tools (Drata, Vanta, Secureframe)\footnote{Attestation-automation tools. Drata: \url{https://drata.com/}; Vanta: \url{https://www.vanta.com/}; Secureframe: \url{https://secureframe.com/}} target SOC\,2 and ISO~27001 and lack CRA technical-file or declaration support; SBOM/SCA scanners and CVE feeds produce findings unmapped to CRA articles, evidence, or owners; and ENISA crosswalks are read-only catalogues \cite{enisa_cra_guidance}. Unlike these tools, CYBERFORT ties every artefact to a specific CRA Annex~I obligation and to a line item of the EU declaration of conformity, integrating standards mapping, SBOM, risk, controls, evidence, and conformity in a single compliance chain aligned with the NIST~CSF~2.0 functions~\cite{nist_csf20}.

\section{CYBERFORT Platform Overview}\label{sec:overview}
\begin{figure}[t]
\centering
\includegraphics[width=0.86\columnwidth]{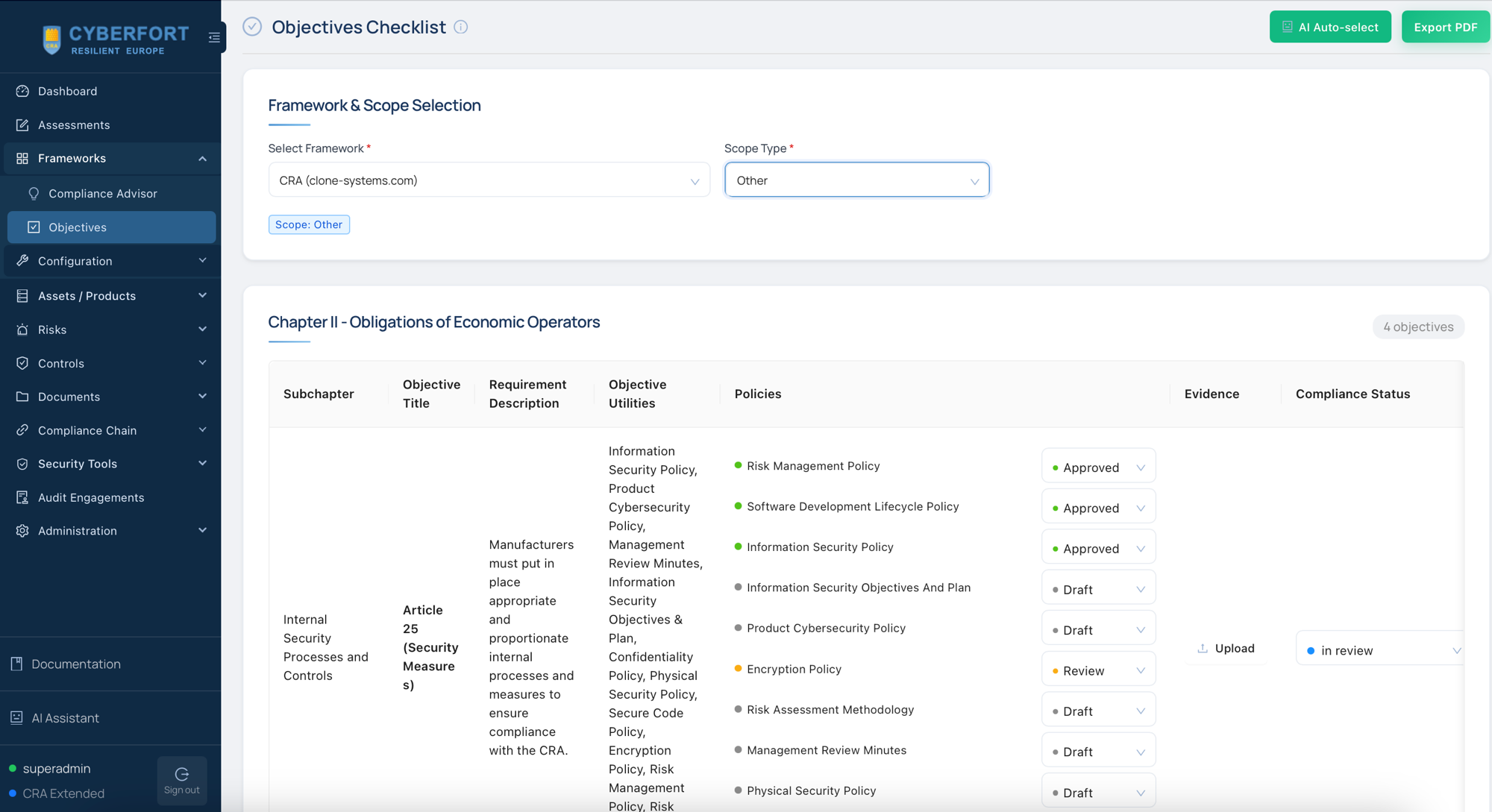}
\caption{CYBERFORT objectives checklist: a CRA Chapter~II economic-operator obligation linked to policies, evidence, and compliance status, i.e.\ the compliance chain in the UI, with the CRA-oriented navigation.}
\label{fig:objectives}
\end{figure}

CYBERFORT is a web-based GRC platform built on a FastAPI backend, a React front end, and a PostgreSQL database, with role-based access control, framework seed data, and integrations to containerised security-assessment tools. Its data model treats organisations, users, frameworks, objectives, assets, products, risks, controls, policies, assessments, and documents as first-class entities. The live deployment (Fig.~\ref{fig:objectives}) exposes a left-side navigation that groups the CRA-relevant capabilities: Assessments, Frameworks, Assets/Products, Risks, Controls, Documents (policies, architecture, evidence, EU declaration of conformity), Technical File, Compliance Chain, Security Tools, and Audit Engagements, with headline counters reporting the current first-cohort load (Table~\ref{tab:results}).

\subsection{High-Level Architecture}\label{sec:arch}
Figure~\ref{fig:arch} renders the architecture in five layers: an \emph{input layer} (SME profile, products, assets, scan results); a \emph{framework layer} (CRA as primary framework, with ISO/IEC~27001, NIS2, and GDPR as reusable control sources, GDPR only where personal-data handling intersects CRA); a \emph{service layer} of sixteen modules in four rows (governance, compliance-chain data, operations, automation); an \emph{integration layer} (the \emph{Wrappers} bus exposing OWASP~ZAP, Semgrep, Syft, OSV~Scanner, Nmap as Evidence producers); and an \emph{output layer} (technical documents, risk and gap reports, audit pack, EU declaration draft). Every output traces back to a product, requirement, risk, control, policy, and evidence object.

\input{cyberfort_arch_v5.tex}

\subsection{The Compliance Chain}\label{sec:compliance-chain}
Traditional GRC tools store requirements, controls, policies, and evidence as separate artifacts. The compliance chain instead treats them as connected nodes in a traceable path. The platform stores this path from a product to its CRA obligations and the conformity artefacts:

\begin{quote}
product/asset $\rightarrow$ product risk $\rightarrow$ mapped control $\rightarrow$ policy/procedure $\rightarrow$ CRA obligation (Annex~I objective) $\rightarrow$ assessment answer with linked evidence $\rightarrow$ gap or conformity status $\rightarrow$ corrective action $\rightarrow$ technical documentation $\rightarrow$ EU declaration of conformity.
\end{quote}

A concrete instance from the NextGen SIEM Pro case study (Section~\ref{sec:eval}) traverses the same path end to end:

\begin{quote}\footnotesize
product SBOM component \emph{(vulnerable third-party dependency)} $\rightarrow$ product risk \emph{``known-vulnerable dependency shipped''} $\rightarrow$ control \emph{ISO/IEC~27001 A.8.8, technical-vulnerability management} $\rightarrow$ \emph{vulnerability- and patch-management policy} $\rightarrow$ CRA obligation \emph{Annex~I Part~I \S(2)(a), no known exploitable vulnerabilities} $\rightarrow$ assessment answer \emph{partially covered}, evidenced by an \emph{OSV-Scanner report over the SBOM} $\rightarrow$ \emph{open corrective action (upgrade pending)} $\rightarrow$ technical-file vulnerability-handling record $\rightarrow$ EU declaration line item.
\end{quote}

The chain lets SMEs answer audit questions directly: which CRA requirement a control satisfies, which product risk justifies it, what evidence proves implementation, and which gap blocks readiness. It also supports cross-framework evidence reuse: a single penetration-test report can support CRA secure-by-design, ISO/IEC~27001 vulnerability-management, and NIS2 risk-management obligations, so the platform stores one evidence object linked to many requirements rather than duplicating uploads.

\subsection{Implementation and Engineering}\label{sec:impl}
\paragraph{Chain representation} The compliance chain is stored relationally in PostgreSQL rather than in a separate graph database. Each chain node type (product, asset, risk, control, policy, objective, evidence, assessment answer, corrective action, document) is a first-class table, and each chain edge is a typed association row (for example asset--risk, risk--control, control--policy, policy--objective) carrying the mapping metadata used by gap analysis (mapping type, coverage level, owner, review status, evidence type). A chain query is therefore a parameterised join, expressed as recursive common table expressions over the association tables, returning every node and edge on the product-to-declaration path. A relational design is appropriate because the chain has bounded, well-typed depth and the dominant access pattern is typed traversal and roll-up rather than open-ended path search; it also lets an SME self-host one database engine instead of a dedicated graph store.

\paragraph{API and backend} The backend is layered as routers, services, repositories, and models, with typed data-transfer objects validating every request and response and an auto-generated OpenAPI specification documenting the REST surface. Role-based access-control middleware enforces per-tenant isolation, so a single deployment serves many SMEs. Scan execution is asynchronous: the Orchestrator enqueues scan jobs through the \emph{Wrappers} bus and the API returns immediately, so long-running scans never block the request path.

\paragraph{Deployment and scale} The platform and its scanner wrappers are packaged as containers and deployed with Docker Compose behind a reverse proxy, so an SME can self-host on a single virtual machine or use the hosted instance at \texttt{access.cyber-fort.eu}. That instance sustains the first cohort of 43 organisations on a modest single node; the API workers are stateless and scale horizontally, and because scans run out of band the request path is decoupled from scan cost. Systematic load benchmarking (concurrent assessments, large-SBOM scans, and chain-traversal latency as the graph grows) is planned.

\section{Proposed CRA Modules}\label{sec:cra-modules}
\subsection{Compliance Mapping and Gap Analysis}
The compliance-mapping module is the core of the multi-framework approach \cite{eu_cra_2024,nist_csf20}. It maintains mappings between CRA provisions and controls from other frameworks: CRA secure-by-default requirements may map to configuration-management, secure-development, and vulnerability-management controls in ISO/IEC~27001, and NIS2 incident-handling and vulnerability-disclosure obligations to CRA vulnerability handling and reporting. Mappings are structured objects (source requirement, target control, mapping type, coverage level, evidence type, owner, review status), and each requirement is classified as fully, partially, or not covered, not applicable, or requiring review. Gap analysis then computes missing controls, missing evidence, outdated policies, open corrective actions, and unassessed products.

\subsection{Risk Assessment}
The CRA explicitly requires cybersecurity risk assessment for products with digital elements \cite{eu_cra_2024}. CYBERFORT's risk module links product assets, threats, vulnerabilities, impacts, likelihood, severity, residual risk, and treatment actions, supporting both organisational and product-specific risks; product records additionally capture intended purpose, foreseeable use, operational environment, support period, components, and third-party dependencies. Risk outputs feed the compliance chain: a risk of shipping known exploitable vulnerabilities, for instance, can require a vulnerability-management control, a patch-management policy, scan evidence, and a corrective action.

\subsection{Control and Policy Management}
The controls and policies module maintains the operational implementation layer required by CRA lifecycle governance and technical-documentation obligations \cite{eu_cra_2024}: controls link to risks and requirements, policies document how controls are implemented, and evidence demonstrates implementation. Representative policies span product cybersecurity, secure software development lifecycle, vulnerability and patch management, product lifecycle support, risk-management methodology, logging and monitoring, and market-surveillance communication. Beyond templates that reduce drafting effort, the module tracks ownership, approval, version history, review dates, applicability, mapped requirements, linked evidence, and open exceptions, so that each policy is linked to CRA obligations and product risk rather than merely stored.

\subsection{EU Conformity Support}
The EU conformity module supports the technical documentation and declaration material required before placing products on the EU market \cite{eu_cra_2024}. It collects product information, operator role, criticality class, applicable requirements, conformity route, standards, risk assessment, test evidence, vulnerability-handling procedures, user instructions, and declaration content, and generates a readiness package (technical-documentation checklist, declaration draft, evidence index, open gaps, corrective-action plan). It does not replace notified-body or legal review; it provides structured preparation and traceability so SMEs can see what is missing before formal assessment.

\subsection{Scanner Integration and Evidence Automation}\label{sec:scanners}
\begin{figure}[t]
\centering
\includegraphics[width=0.9\columnwidth]{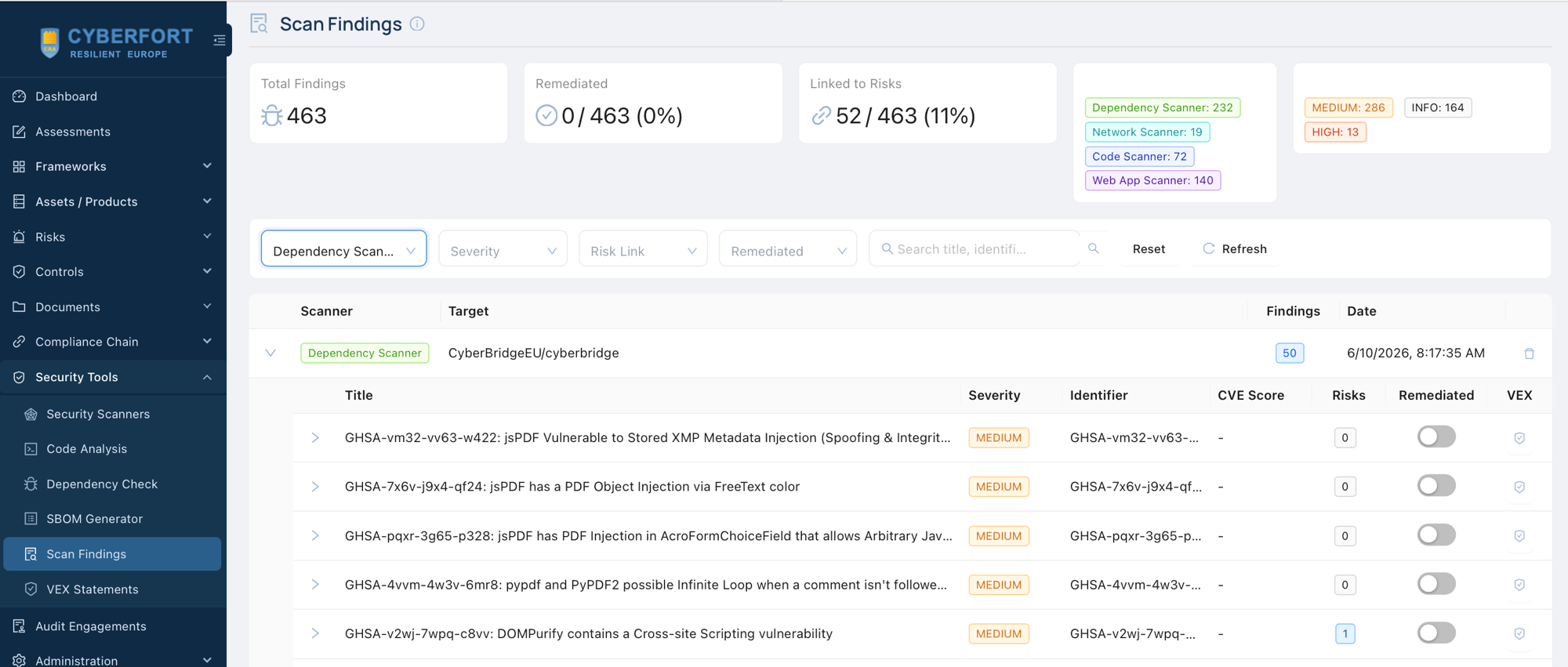}
\caption{Aggregate Scan Findings view: multi-scanner evidence normalised into one register, rolled up by severity and risk linkage, available as Evidence objects for Annex~I Part~II.}
\label{fig:scanfindings}
\end{figure}

The scanner-integration module realises the \emph{Wrappers} layer of Fig.~\ref{fig:arch}: it embeds containerised security-assessment tools (dynamic and static application testing, software-composition analysis, and network discovery) into the compliance chain so that technical findings become first-class evidence. Each wrapper is a microservice with a uniform REST interface, so additional tools can be added without changing the platform core. Each scan execution is anchored to a product, an assessment, and one or more CRA objectives; the Orchestrator schedules scans, normalises results, and creates evidence objects that the mapping and conformity modules attribute to specific Annex~I requirements and product risks. Evidence records retain provenance (tool, version, scan date, target hash, signed result) so auditors can verify freshness against the CRA support period. Figure~\ref{fig:scanfindings} shows the aggregate Scan Findings view for one onboarded product: 463 findings across dependency (232), web-application (140), code (72), and network (19) scanners, rolled up by severity (13 high, 286 medium, 164 informational) and by risk linkage (52, i.e.\ 11\%, already tied to product risks), with each finding available as an Evidence object for the compliance chain.

\subsection{End-to-End Workflow}\label{sec:workflow}
The five modules are run as a six-step workflow that embeds risk assessment across the product lifecycle~\cite{eu_cra_2024}: (i) import the CRA framework template; (ii) register products and identify operator role and criticality; (iii) perform a product risk assessment; (iv) run a conformity assessment, linking answers to evidence and policies; (v) compute coverage and gaps and assign corrective actions; and (vi) export a readiness package.

\section{Evaluation}\label{sec:eval}
We evaluate whether CYBERFORT reduces CRA preparation effort and improves traceability for SMEs. The evaluation reuses pilot scenarios from CYBERFORT deliverables rather than synthetic requirements, and spans four SME profiles covering the economic-operator roles and product types SMEs most commonly encounter: (i) a managed-security-services manufacturer (Clone Systems) of \textit{NextGen SIEM Pro} (\textit{SEUXDR}), a Wazuh~4.x and OpenSearch SIEM/XDR\footnote{Wazuh: \url{https://wazuh.com/}; OpenSearch: \url{https://opensearch.org/}} with a ReAct-based~\cite{yao2023react} active-response agent, classified as an Annex~III Class~I important product; (ii) a regulated fintech (Dealio) for continuous compliance management; (iii) a maritime operator (Columbia) for shipboard and operational-value-chain preparedness; and (iv) an energy-sector provider (Eximprod Engineering) for vulnerability assessment and remediation reporting on hardware-plus-software products. Only the NextGen SIEM Pro case has been executed end-to-end on the live platform; it yields the measured results below, while the other three pilots are provisioned and in progress.

\subsection{Metrics}
The evaluation records five metrics: \textit{requirement coverage and gap detection} (share of applicable CRA objectives mapped to controls, policies, risks, and evidence, and the number and severity of gaps); \textit{time-to-assessment} (time to complete the first CRA conformity assessment and export a report); \textit{evidence completeness and reuse} (share of answers with linked evidence, and evidence artifacts reused across CRA, ISO/IEC~27001, NIS2, GDPR, and NIST~CSF~2.0); \textit{conformity-document readiness} (completeness of the technical-documentation checklist, evidence index, records, and EU declaration draft); and \textit{user acceptance} (usability, clarity, and confidence in outputs).

\subsection{Measured Results: NextGen SIEM Pro}
The NextGen SIEM Pro scenario was executed end-to-end on the live CYBERFORT instance with Clone Systems acting as the manufacturer. The product was registered as software with digital elements, classified as an Annex~III Class~I important product, and given an attached SBOM; a CRA risk register, control register, policy set, and conformity assessment were then built on the same instance, and Table~\ref{tab:results} reports the measured outcomes against the metrics above. CYBERFORT's CRA framework template decomposes Annex~I (Parts~I and~II) together with the Article-level lifecycle duties into 42 assessable objectives; for this product 27 were mapped to at least one control and the remaining 15 surfaced as gaps. Covered objectives concentrated on application and API security, identity and access control, automation oversight, vulnerability exploitation, supply chain, governance, data security, and operational resilience. The integrated scanner layer was run and its findings ingested as Evidence: an SBOM (Syft) was attached and analysed for known-vulnerable dependencies (OSV-Scanner), with static analysis (Semgrep), dynamic application testing (OWASP~ZAP), and network discovery (Nmap) contributing further findings. Each finding is stored as an Evidence object carrying its provenance (tool, version, scan date, target) and mapped to the relevant Annex~I Part~II vulnerability-handling obligation. Every assessment answer links to at least one control, one policy, and one CRA objective, and every covered objective links back to at least one product risk, giving full intra-chain traceability for the covered scope; the 15 remaining objectives surface as explicit, assignable gaps. The same instance concurrently sustains the first cohort of 43 organisations (Table~\ref{tab:results}), indicating multi-tenant scale rather than a single case.

\begin{table}[t]
\caption{Measured results for the completed NextGen SIEM Pro case study and the live first-cohort deployment}
\label{tab:results}
\centering
\footnotesize
\renewcommand{\arraystretch}{0.95}
\begin{tabular}{p{0.30\columnwidth}p{0.60\columnwidth}}
\toprule
\textbf{Metric} & \textbf{Measured value} \\
\midrule
Requirement coverage & 27 of 42 CRA objectives (64\%) mapped to $\geq$1 control, spanning the CRA obligation chapters \\
Gap detection & 15 uncovered objectives flagged as assignable gaps; 30 risks at severity 12 critical, 14 high, 4 medium \\
Risk selection & 30 product risks narrowed from a 104-item template library \\
Control mapping & 65 controls (55 baseline plus 10 SCF AAT controls for autonomous operation) \\
Policy linkage & 32 policies linked as compliance-chain nodes \\
Automated evidence & Multi-scanner findings (dependency, code, web-application, network) ingested as Evidence objects mapped to Annex~I Part~II \\
Traceability completeness & 100\% of covered answers linked to $\geq$1 control, policy, and objective; every covered objective linked to $\geq$1 risk \\
Conformity-document readiness & Technical-file checklist, evidence index, and EU declaration draft generated \\
Deployment scale (live) & 43 organisations, 109 users, 60 frameworks, 918 policies, 1213 risks \\
\bottomrule
\end{tabular}
\end{table}

\subsection{Threats to Validity and Planned Controlled Study}
The principal limitation is that the completed case study was performed by Clone Systems, the SME that also develops CYBERFORT, so it is an existence proof for end-to-end traceability rather than independent evidence of usability and effectiveness; we accordingly restrict its claims to chain expressiveness for a realistic scenario combining conventional and AI/autonomous-operation risks, and full intra-chain traceability for the covered scope. The three independent pilots (Dealio, Columbia, Eximprod Engineering) are provisioned to supply the external evidence this case cannot.

The metrics that require a baseline (time-to-assessment, cross-framework evidence-reuse magnitude, and user-acceptance ratings) are not reported here because the controlled study is still running. That study pairs each pilot SME against a spreadsheet-and-free-form baseline and adds a structured expert review in which two to four practitioners (SME consultants, internal CISOs, and a notified-body-facing assessor) audit the generated compliance chain, declaration draft, technical-file checklist, and evidence index and rate usability and trustworthiness. Consistent with prior SME compliance work~\cite{conceicao2026sme,szedlak2024cra}, we expect reduced duplicate evidence handling, higher traceable coverage, and shorter time to a readiness package, and report it as an indicative effort study rather than a controlled experiment.

\subsection{Coverage of CRA Obligations}
Table~\ref{tab:coverage} summarises which CRA obligation families are operationally supported by CYBERFORT today and which remain partial or out of scope, making explicit which artifacts the platform produces for each obligation family.

\begin{table}[t]
\caption{CRA obligation coverage by CYBERFORT modules}
\label{tab:coverage}
\centering
\footnotesize
\renewcommand{\arraystretch}{0.95}
\begin{tabular}{p{0.34\columnwidth}p{0.10\columnwidth}p{0.42\columnwidth}}
\toprule
\textbf{CRA obligation family} & \textbf{Support} & \textbf{Primary CYBERFORT artifact} \\
\midrule
Product scoping and economic-operator role (Arts.~3,~13) & Full & Product registry, scope-assessment screen \\
Essential cybersecurity requirements (Annex~I, Part~I) & Full & Framework objectives, conformity self-assessment, control mapping \\
Vulnerability handling (Annex~I, Part~II) & Full & Vulnerability-management policy, SBOM linkage, evidence records \\
Cybersecurity risk assessment (Art.~13) & Full & Product risk register, residual risk, treatment plan \\
Technical documentation (Annex~VII) & Partial & Technical-file checklist, evidence index \\
EU declaration of conformity (Annex~V) & Partial & Declaration draft generator and evidence references \\
Conformity assessment procedures (Art.~32) & Partial & Routing per criticality class; notified-body workflow planned \\
Reporting obligations (Art.~14) & Partial & Vulnerability and incident records; ENISA portal integration planned \\
Market surveillance and CE marking (Art.~30, Ch.~V) & Partial & CE marking checklist; market-surveillance workflow planned \\
Authorised representative, importer, distributor duties (Arts.~18--20) & Partial & Role attributes on product; per-role workflows planned \\
\bottomrule
\end{tabular}
\end{table}

\section{Discussion, Conclusion, and Future Work}
This paper presented CYBERFORT, a CRA-first compliance-chain platform for SMEs, with four CRA-oriented modules and an integrated scanner layer that turns technical findings into Evidence objects linked to CRA Annex~I obligations and an EU declaration of conformity. The engineering account (Section~\ref{sec:impl}) realises the chain on a relational data model with typed association tables, an asynchronous scan orchestrator, and a self-hostable container deployment, while the completed case study (Section~\ref{sec:eval}) demonstrates full intra-chain traceability across 27 of 42 CRA objectives for a SIEM/XDR product spanning conventional and AI/autonomous-operation risks, at multi-tenant scale.

Four directions remain: making cross-mapping evidence-aware, with each mapping object carrying strength, assumptions, evidence types, and review status; extending evidence automation from the scanner wrappers to CI/CD pipelines, ticketing systems, and signed approvals, with freshness measured against the CRA support period; adding role-specific conformity workflows for importers, distributors, and authorised representatives, and integrating the ENISA single-reporting platform for Art.~14; and running the controlled effort study and expert-review audit of Section~\ref{sec:eval} across the three independent pilots to supply the comparative and external evidence the present case study cannot.

\section*{Acknowledgment}
This work has received funding from the European Union's Digital Europe Programme under grant agreement No~101190281 (CYBERFORT). The authors thank the whole CYBERFORT consortium and SME stakeholders whose requirements informed the platform design.

\bibliographystyle{IEEEtran}
\bibliography{references}

\end{document}

%% file: cyberfort_arch_v5.tex

\begin{figure*}[t]
\centering
\resizebox{\textwidth}{!}{%
\begin{tikzpicture}[
  font=\footnotesize,
  cra/.style    ={rectangle,rounded corners=2pt,draw,thick,fill=blue!22,
                  minimum height=7mm,minimum width=22mm,align=center,inner sep=2pt,
                  font=\footnotesize\bfseries},
  frmw/.style   ={rectangle,rounded corners=2pt,draw,fill=gray!12,
                  minimum height=6mm,minimum width=17mm,align=center,inner sep=1.5pt},
  ustep/.style  ={rectangle,rounded corners=2pt,draw,fill=orange!20,
                  minimum height=5mm,minimum width=34mm,align=center,inner sep=1.5pt},
  pilot/.style  ={rectangle,rounded corners=2pt,draw,fill=orange!20,
                  minimum height=5mm,minimum width=17mm,align=center,inner sep=1.5pt},
  module/.style ={rectangle,rounded corners=2pt,draw,fill=gray!18,
                  minimum height=6mm,minimum width=20mm,align=center,inner sep=1.5pt},
  wrap/.style   ={rectangle,rounded corners=2pt,draw,fill=gray!25,
                  minimum height=5mm,align=center,inner sep=2pt},
  tool/.style   ={rectangle,rounded corners=2pt,draw,fill=gray!15,
                  minimum height=6mm,minimum width=15mm,align=center,inner sep=1.5pt,
                  font=\scriptsize},
  outbox/.style ={rectangle,rounded corners=2pt,draw,fill=yellow!25,
                  minimum height=5mm,minimum width=33mm,align=center,inner sep=1.5pt},
  fb/.style     ={rectangle,rounded corners=2pt,draw,fill=yellow!25,
                  minimum height=8mm,minimum width=33mm,align=center,inner sep=1.5pt},
  grp/.style    ={rectangle,rounded corners=3pt,draw,thick,fill=none,inner sep=2.5mm},
  arrow/.style  ={-Latex,thick,gray!75},
  fbarrow/.style={-Latex,thick,red!60!black,dashed},
]

\node[cra,minimum width=26mm] (fwCRA) at (-1.0, 8.0) {EU CRA};

\node[frmw,minimum width=13mm] (fw2) at (1.6, 8.0) {ISO/IEC\\27001};
\node[frmw,minimum width=13mm] (fw5) at (3.2, 8.0) {NIS2};
\node[frmw,minimum width=13mm] (fw7) at (4.8, 8.0) {GDPR};

\node[ustep] (s1) at (-7.5, 5.6) {Enter name};
\node[ustep] (s2) at (-7.5, 4.8) {Enter Organization};
\node[ustep] (s3) at (-7.5, 4.0) {Define assets};
\node[ustep] (s4) at (-7.5, 3.2) {Choose frameworks};
\node[ustep] (s5) at (-7.5, 2.4) {Check scope, requirements};
\node[ustep] (s6) at (-7.5, 1.6) {Maturity};
\foreach \a/\b in {s1/s2,s2/s3,s3/s4,s4/s5,s5/s6} \draw[arrow] (\a) -- (\b);

\node[pilot] (p1) at (-8.4, 0.0) {Maritime};
\node[pilot] (p2) at (-6.6, 0.0) {Energy};
\node[pilot] (p3) at (-8.4,-0.8) {Finance};
\node[pilot] (p4) at (-6.6,-0.8) {SOC / MSSP};

\node[module] (m11) at (-2.5, 5.6) {Assessments};
\node[module] (m12) at ( 0.0, 5.6) {Documents};
\node[module] (m13) at ( 2.5, 5.6) {Policies};
\node[module] (m14) at ( 5.0, 5.6) {Compliance\\Advisory};

\node[module] (m21) at (-2.5, 4.6) {Risks};
\node[module] (m22) at ( 0.0, 4.6) {Evidences};
\node[module] (m23) at ( 2.5, 4.6) {Objectives};
\node[module] (m24) at ( 5.0, 4.6) {Report\\Generator};

\node[module] (m31) at (-2.5, 3.6) {Controls};
\node[module] (m32) at ( 0.0, 3.6) {Incidents};
\node[module] (m33) at ( 2.5, 3.6) {Auditor\\Assignment};
\node[module] (m34) at ( 5.0, 3.6) {Scope\\Assessment};

\node[module] (m41) at (-2.5, 2.6) {Orchestrator};
\node[module] (m42) at ( 0.0, 2.6) {AI Engine};
\node[module] (m43) at ( 2.5, 2.6) {Compliance\\Mapping};
\node[module] (m44) at ( 5.0, 2.6) {Gap\\Analysis};

\node[wrap,minimum width=112mm] (wrap) at (1.25, 1.55) {Wrappers (integration bus)};

\node[tool] (t1) at (-2.55,-0.4) {DAST\\\textit{(dynamic)}};
\node[tool] (t2) at (-0.65,-0.4) {SAST\\\textit{(static)}};
\node[tool] (t3) at ( 1.25,-0.4) {SBOM\\generation};
\node[tool] (t4) at ( 3.15,-0.4) {SCA \&\\vuln. scan};
\node[tool] (t5) at ( 5.05,-0.4) {Network\\discovery};

\node[outbox] (o1) at (10.0, 5.6) {Technical Documents};
\node[outbox] (o2) at (10.0, 4.8) {Risk Results};
\node[outbox] (o3) at (10.0, 4.0) {Declaration of Conformity};
\node[outbox] (o4) at (10.0, 3.2) {Audit Pack};
\node[outbox] (o5) at (10.0, 2.4) {Assessment Results};
\node[outbox] (o6) at (10.0, 1.6) {Artifacts (PDF / Word / Excel)};

\node[fb] (fbk) at (10.0,-0.4) {SMEs Feedback /\\Content Improvement};

\node[grp,fit=(fwCRA),
      label={[font=\scriptsize\bfseries,blue!50!black]above:Primary framework}] (fwsCRA) {};
\node[grp,fit=(fw2)(fw7),
      label={[font=\scriptsize\bfseries]above:Additional reusable frameworks}] (fwsADD) {};
\node[grp,fit=(s1)(s6),
      label={[font=\small\bfseries]above:User / SME}] (user) {};
\node[grp,fit=(p1)(p4),
      label={[font=\small\bfseries]above:Pilots}] (pilots) {};
\node[grp,fit=(m11)(m44)(wrap),
      label={[font=\small\bfseries]above:CYBERFORT Platform}] (plat) {};
\node[grp,fit=(t1)(t5),
      label={[font=\small\bfseries]above:Assessment tools (containerized)}] (tools) {};
\node[grp,fit=(o1)(o6),
      label={[font=\small\bfseries]above:Outputs}] (outs) {};

\draw[arrow] (fwsCRA.south) -- (fwsCRA.south |- plat.north);
\draw[arrow] (fwsADD.south) -- (fwsADD.south |- plat.north);

\draw[arrow] (user.east) -- node[midway,fill=white,inner sep=1.5pt,font=\scriptsize,align=center]{onboarding\\inputs} (plat.west);

\draw[arrow] (pilots.east) -- (plat.south west);

\draw[arrow] (plat.east) -- node[midway,fill=white,inner sep=1.5pt,font=\scriptsize,align=center]{traceable\\outputs} (outs.west);

\draw[arrow] (wrap.south) -- (tools.north);

\draw[arrow,dashed] (tools.east) to[bend right=20]
    node[below,font=\scriptsize,sloped]{scan evidence} (plat.south east);

\draw[fbarrow] (outs.south) -- (fbk.north);
\draw[fbarrow] (fbk.north west) to[out=120,in=-20]
    node[midway,above,font=\scriptsize,sloped]{feedback}
    (plat.south east);

\end{tikzpicture}%
}
\caption{CYBERFORT architecture. The regulatory layer (top) splits the primary \emph{EU CRA} framework from reusable frameworks (ISO/IEC~27001, NIS2, GDPR). The central platform exposes sixteen modules in four rows (governance, compliance-chain data, operations, automation) over a \emph{Wrappers} bus connecting the containerised assessment tools. SME inputs and pilots sit left, CRA-oriented outputs right, with a dashed feedback loop.}
\label{fig:arch}
\end{figure*}

%% file: references.bib
@misc{eu_cra_2024,
  author={{European Parliament and Council of the European Union}},
  title={{Regulation (EU) 2024/2847 on horizontal cybersecurity requirements for products with digital elements (Cyber Resilience Act)}},
  year={2024},
  howpublished={Official Journal of the European Union},
  url={https://eur-lex.europa.eu/eli/reg/2024/2847/oj/eng}
}

@misc{ec_cra_policy,
  author={{European Commission}},
  title={{Cyber Resilience Act}},
  year={2025},
  url={https://digital-strategy.ec.europa.eu/en/policies/cyber-resilience-act}
}

@misc{enisa_sme_cra,
  author={{ENISA}},
  title={{SME Cyber Resilience Act Survey}},
  year={2025},
  url={https://www.enisa.europa.eu/topics/awareness-and-cyber-hygiene/sme-cyber-resilience-act-survey}
}

@inproceedings{szedlak2024cra,
  author={Szedlak, Christoph and Reinemann, Holger and Hatzelmann, Sophia},
  title={{Ensuring Cybersecurity Compliance: Assessing SME Awareness and Preparedness for the Cyber Resilience Act}},
  booktitle={Proceedings of the 5th Asia Pacific Conference on Industrial Engineering and Operations Management},
  year={2024},
  doi={10.46254/AP05.20240160}
}

@inproceedings{yao2023react,
  author={Yao, Shunyu and Zhao, Jeffrey and Yu, Dian and Du, Nan and Shafran, Izhak and Narasimhan, Karthik and Cao, Yuan},
  title={{ReAct: Synergizing Reasoning and Acting in Language Models}},
  booktitle={Proceedings of the International Conference on Learning Representations (ICLR)},
  year={2023}
}

@misc{semgrep,
  author={{Semgrep Inc.}},
  title={{Semgrep: Static analysis at ludicrous speed}},
  year={2026},
  url={https://semgrep.dev/}
}

@misc{syft,
  author={{Anchore}},
  title={{Syft: SBOM generation tool}},
  year={2026},
  url={https://github.com/anchore/syft}
}

@misc{scf_aat,
  author={{Secure Controls Framework}},
  title={{SCF v2025.1: AI \& Autonomous Technologies (AAT) family}},
  year={2025},
  url={https://securecontrolsframework.com/}
}

@misc{nist_csf20,
  author={{National Institute of Standards and Technology}},
  title={{The NIST Cybersecurity Framework (CSF) 2.0}},
  year={2024},
  url={https://www.nist.gov/cyberframework}
}

@article{teichmann2025clsr,
  author={Teichmann, Fabian and Sergi, Bruno S.},
  title={{The EU Cyber Resilience Act: Hybrid governance, compliance, and cybersecurity regulation in the digital ecosystem}},
  journal={Computer Law \& Security Review},
  volume={59},
  pages={106209},
  year={2025},
  doi={10.1016/j.clsr.2025.106209}
}

@article{conceicao2026sme,
  author={Conceicao, Francisco and Rocha, Manuel and Almeida, Fernando},
  title={{Security Compliance as a Catalyst for Sustainable Partnerships: A Design Science Approach for SMEs}},
  journal={Journal of Cybersecurity and Privacy},
  volume={6},
  number={2},
  pages={53},
  year={2026},
  doi={10.3390/jcp6020053}
}

@misc{enisa_cra_guidance,
  author={{ENISA}},
  title={{Cyber Resilience Act Requirements Standards Mapping}},
  year={2024},
  url={https://www.enisa.europa.eu/publications/cyber-resilience-act-requirements-standards-mapping}
}

@misc{ec_cra_msme,
  author={{European Commission}},
  title={{The Cyber Resilience Act and MSMEs: EU-funded projects supporting SME compliance}},
  year={2025},
  howpublished={Shaping Europe's digital future},
  url={https://digital-strategy.ec.europa.eu/en/policies/cra-msmes}
}
